\documentclass[aps,twocolumn,floats,prd,psfig]{revtex4}
\usepackage{graphicx, epsfig}

\newcommand{\PRL}{Phys. Rev. Lett.}
\newcommand{\PRD}{Phys. Rev. D}

\newcommand{\beq}{\begin{equation}}
\newcommand{\eeq}{\end{equation}}
\newcommand{\beqa}{\begin{eqnarray}}
\newcommand{\eeqa}{\end{eqnarray}}

\newcommand {\ds}{\displaystyle}

\begin{document}

\title{Uncorrelated Estimates of Dark Energy Evolution}
\author{Dragan Huterer}  
\affiliation{Department of Physics, Case Western Reserve University,
Cleveland, OH 44106; dhuterer@cwru.edu \vspace{0.0cm}}

\author{Asantha Cooray}
\affiliation{Theoretical Astrophysics, Division of Physics,
Mathematics and Astronomy, California Institute of Technology, MS
130-33, Pasadena, CA 91125; asante@caltech.edu} 


\begin{abstract}
Type Ia supernova data have recently become strong enough to enable,
for the first time, constraints on the time variation of the dark
energy density and its equation of state. Most analyses, however, are
using simple two or three-parameter descriptions of the dark energy
evolution, since it is well known that allowing more degrees of
freedom introduces serious degeneracies. Here we present a method to
produce uncorrelated and nearly model-independent band power estimates
of the equation of state of dark energy and its density as a function
of redshift. We apply the method to recently compiled supernova
data.  Our results are consistent with the cosmological constant
scenario, in agreement with other analyses that use traditional
parameterizations, though we find marginal (2-$\sigma$) evidence for
$w(z)<-1$ at $z<0.2$. In addition to easy interpretation,
uncorrelated, localized band powers allow intuitive and powerful
testing of the constancy of either the energy density or equation of
state.  While we have used relatively coarse redshift binning suitable
for the current set of $\sim$ 150 supernovae, this approach should
reach its full potential in the future, when applied to thousands of
supernovae found from ground and space, combined with complementary
information from other cosmological probes.
\end{abstract} 

\maketitle


\section{Introduction}
Recent measurements of the distance-redshift relation using type Ia
supernovae (SNe Ia) \cite{Rieetal04,Baretal04} obtained
using the Hubble Space Telescope further strengthened the evidence
that the rate of expansion of the universe is increasing in time
\cite{perlmutter-1999}. This accelerated expansion is
ascribed to a mysterious component called dark energy that comprises
about 70\% of the energy density of the universe. In addition to
supernova data, additional pieces of evidence come from the combined
study of the large scale structure and the cosmic microwave background
anisotropy measurements \cite{Tegetal03}.  While the presence of
dark energy is by now well established, we are at an early stage
of studying and understanding this component. It is hoped that more
accurate cosmological measurements will further constrain 
parameters describing dark energy and eventually shed light on the
underlying physical mechanism.

Dark energy is most simply described by its present day energy density
relative to the critical value, $\Omega_{DE}$, and its equation of state defined
as the ratio of pressure to density, $w\equiv p_{DE}/\rho_{\rm
DE}$ \cite{turner_white}. In general, $w$ is allowed to freely vary with
time (or redshift), as is $\rho_{DE}$. In practice, it is difficult to
constrain $w(z)$ or, say, the scaled energy density $f(z)\equiv
\rho_{DE}(z)/\rho_{DE}(0)$, when they are described by more than a few
parameters due to severe parameter degeneracies entering the
observable quantity (luminosity distance, in the case of SNe Ia). Even
though it is in principle possible to recover the function $f(z)$ or
$w(z)$ directly from supernova measurements
\cite{Star98}, in practice one has to fit the noisy
data with a smooth functional form \cite{HutTur01} which
introduces error and bias (for a valiant attempt to do this with
current data, see \cite{DalDjo04}).  Another general approach is to
model $w$ or $f$ using a cubic spline in redshift
(e.g. \cite{WanTeg04}), but again the paucity of data limits the
spline to a few points in redshift, while having more points would correlate the
measurements making the interpretation somewhat difficult.

Constraints from the new SN Ia data~\cite{Rieetal04} suggest that dark
energy is consistent with the cosmological constant scenario
\cite{Rieetal04,WanTeg04}, agreeing with previous work
\cite{Alaetal03,WanFre04}.  However, these (and other) analyses are
typically based on particular models --- either a linear variation
with redshift \cite{CooHut99} or the evolution that asymptotes to a
constant $w$ at high redshift \cite{Lin03}, or perhaps a more
complicated parameterization \cite{CorCop02} --- that are
used to describe redshift variation of the dark energy equation of
state.  While these forms do a very good job in fitting $w(z)$
due to a variety of proposed mechanisms that could be responsible for
dark energy \cite{Lin04}, one should keep in mind that we are far from
having any solid leads as to what to expect for the dark energy
evolution. Given the constant increase in the quality and quantity of
SN Ia data, it is timely to consider whether one can use current data to
derive model independent conclusions on the evolution of dark energy.

In this paper, we introduce a variant of the principal component
analysis advocated in Ref.~\cite{HutSta03}. We make use of the most recent type Ia
supernova data from Ref.~\cite{Rieetal04} and present a view of dark
energy complementary to other approaches. At the same time, we are
seeking to answer one of the most important questions at present: is
dark energy consistent with the cosmological constant scenario or not?
Our analysis is facilitated by the fact that our measurements are
completely uncorrelated. Finally, we briefly comment on the applicability of
this approach to future datasets. Throughout we assume a flat universe.
 
\section{Methodology}\label{sec:method}

We would like to impose constraints on the parameters $p_i$
($i=1\ldots N$) that describe the dark energy equation of state $w(z)$
or its energy density $f(z)$, each $p_i$ being suitably defined in the
$i^{\rm th}$ redshift bin. In addition to these, we have two more
parameters: matter density relative to the critical 
$\Omega_M=1-\Omega_{DE}$ and the Hubble constant $h\equiv
H_0/$(100km/s/Mpc).  We first marginalize the full
$(N+2)$--dimensional likelihood over these two (for the priors and
assumptions, see Sec.~\ref{sec:results}), and project them onto the
$p_i$ space. The covariance of the $N$ resulting parameters is
\begin{equation}
{\bf C} \equiv \langle {\bf p p^T}\rangle - 
\langle {\bf p}\rangle \langle {\bf p}^T \rangle
\label{eq:cov}
\end{equation}
\noindent where ${\bf p}$ is the vector of parameters $p_i$ and ${\bf p}^T$
its transpose. These parameters can now be rotated into a basis where
they are diagonal by choosing an orthogonal matrix {\bf W} so that
it diagonalizes the Fisher matrix
\begin{equation}
{\bf F}\equiv {\bf C}^{-1} = {\bf W}^T {\bf \Lambda} {\bf W}
\end{equation}
\noindent where ${\bf \Lambda}$ is diagonal. It is clear that the new
parameters $q_i$, defined as ${\bf q}\equiv {\bf Wp}$, are
uncorrelated, for they have the covariance matrix ${\bf
\Lambda}^{-1}$. The $q_i$ are referred to as the principal components
and the rows of ${\bf W}$ are the window functions (or weights) that define how the
principal components are related to the $p_i$. We refer the reader to
Huterer \& Starkman \cite{HutSta03} for a discussion on the
application of principal components to the dark energy equation of
state.

Let us now define ${\bf \widetilde{W}}$ by absorbing the diagonal
elements of ${\bf \Lambda}^{1/2}$ into the corresponding rows of ${\bf
W}$, so that ${\bf \widetilde{W}}^T {\bf \widetilde{W}}={\bf F}$
. Then, as emphasized by Hamilton and Tegmark \cite{HamTeg00} in the
context of matter power spectrum measurements, there are infinitely
many choices for the matrix ${\bf \widetilde{W}}$, as for any
orthogonal matrix ${\bf O}$, ${\bf O\widetilde{W}}$ is also a valid
choice that makes the parameters $q_i$ uncorrelated.  While the
principal components, $q_i$, have several nice features --- in
particular, the best-determined $q_i$ are smoother and have support at
lower redshift than the poorly determined ones --- their corresponding
window functions are oscillatory, making the intuitive interpretation
of the components somewhat difficult.

Here we advocate another choice for the weight matrix ${\bf
\widetilde{W}}$: the square root of the Fisher matrix, ${\bf
\widetilde{W}}={\bf F}^{1/2}\equiv {\bf C}^{-1/2}$
\cite{HamTeg00}. This choice is interesting since the weights
(rows of ${\bf \widetilde{W}}$) are almost everywhere positive, with
very small negative contributions, and this has been recognized as a
useful basis in which to represent measurements of the galaxy power
spectrum from large-scale structure surveys \cite{teg_SDSS_3DPS}. The
matrix ${\bf \widetilde{W}}$ is computed by first diagonalizing the
Fisher (inverse covariance) matrix, ${\bf F}={\bf O}^T {\bf \Lambda
O}$, and then defining ${\bf \widetilde{W}}={\bf O}^T {\bf
\Lambda^{1/2} O}$. We normalize ${\bf \widetilde{W}}$ so that its
rows, the weights for $p_i$, sum to unity. With this choice,
Eq.~(\ref{eq:cov}) shows that the covariance of the new parameters,
${\bf q}\equiv {\bf \widetilde{W}p}$, is
\begin{equation}
\langle \,(q_i-\langle q_i\rangle)\,(q_j-\langle q_j\rangle)\,\rangle=
{\delta_{ij}\over \ds\sum_a (F^{1/2})_{ia} \sum_b (F^{1/2})_{jb}} \, ,
\end{equation}
\noindent 
and parameters $q_i$ are manifestly uncorrelated. Furthermore, their
weights are mostly positive and are localized in redshift fairly
well. We illustrate this in the next section using current supernova
data.

\begin{figure*}[t]
\centerline{\psfig{file=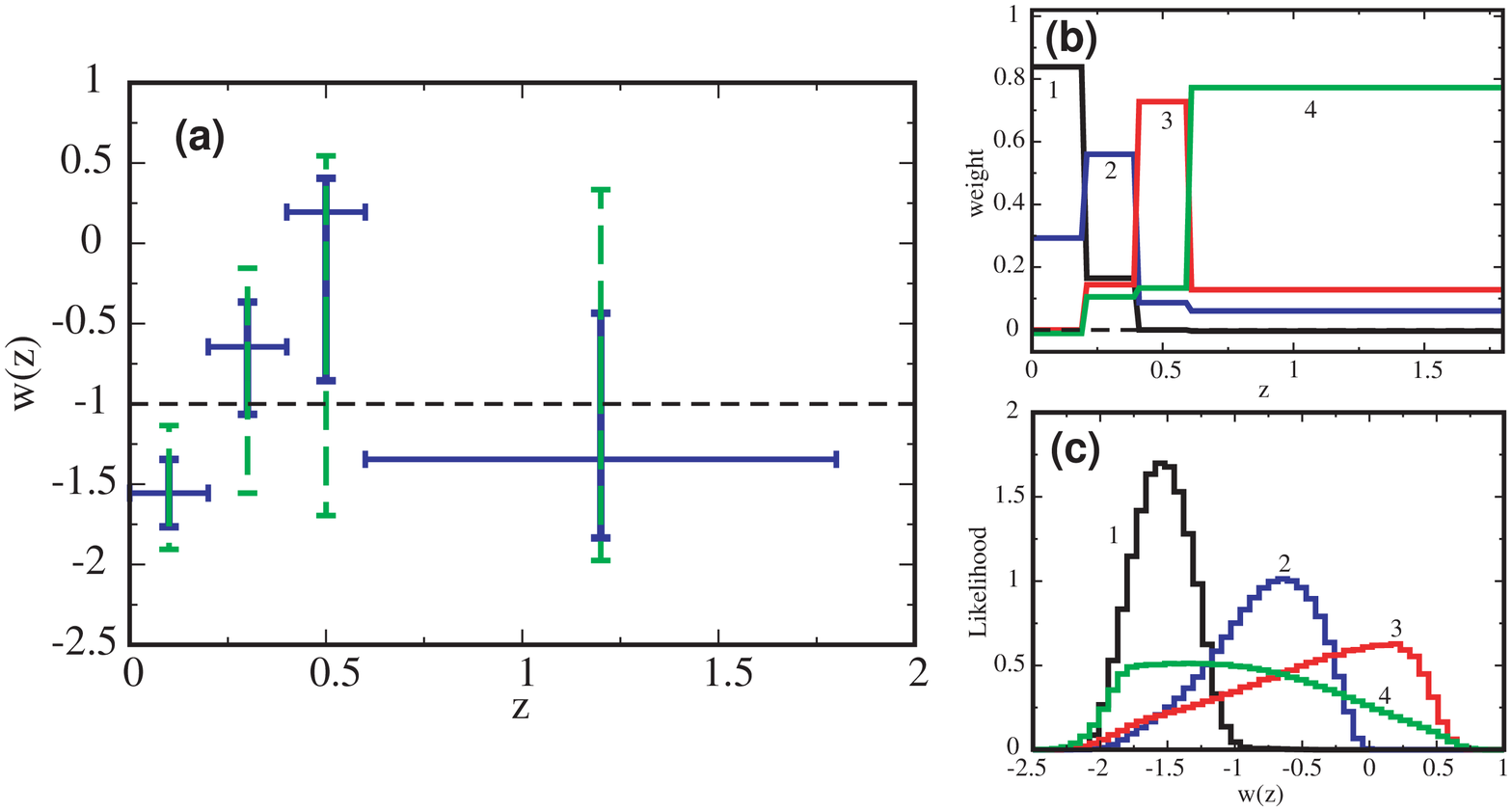,width=6.0in,angle=0}}
\caption{Uncorrelated band-power estimates of the equation of state
$w(z)$ of dark energy are shown in panel (a). Vertical error bars show
the 1 and 2-$\sigma$ error bars (the full likelihoods are shown in
panel (c)), while the horizontal error bars represent the approximate
range over which each measurement applies. The full window functions in
redshift space for each of these measurements are shown in panel (b);
they have small leakage outside of the original redshift divisions.
The window functions and the likelihoods are labeled in order of 
increasing redshift of the band powers in panel (a).
The window functions satisfy three of our requirements: they make the
band-powers uncorrelated, they are fairly well localized in redshift,
and they are almost everywhere positive. In panel (a), we have used
a uniform prior of $0.22\leq \Omega_M \leq 0.38$ (a Gaussian prior
$\Omega_M=0.3\pm 0.04$ gives very similar results), and we
have assumed a flat universe throughout.}
\label{fig:wz}
\end{figure*}

\begin{figure*}[t]
\centerline{\psfig{file=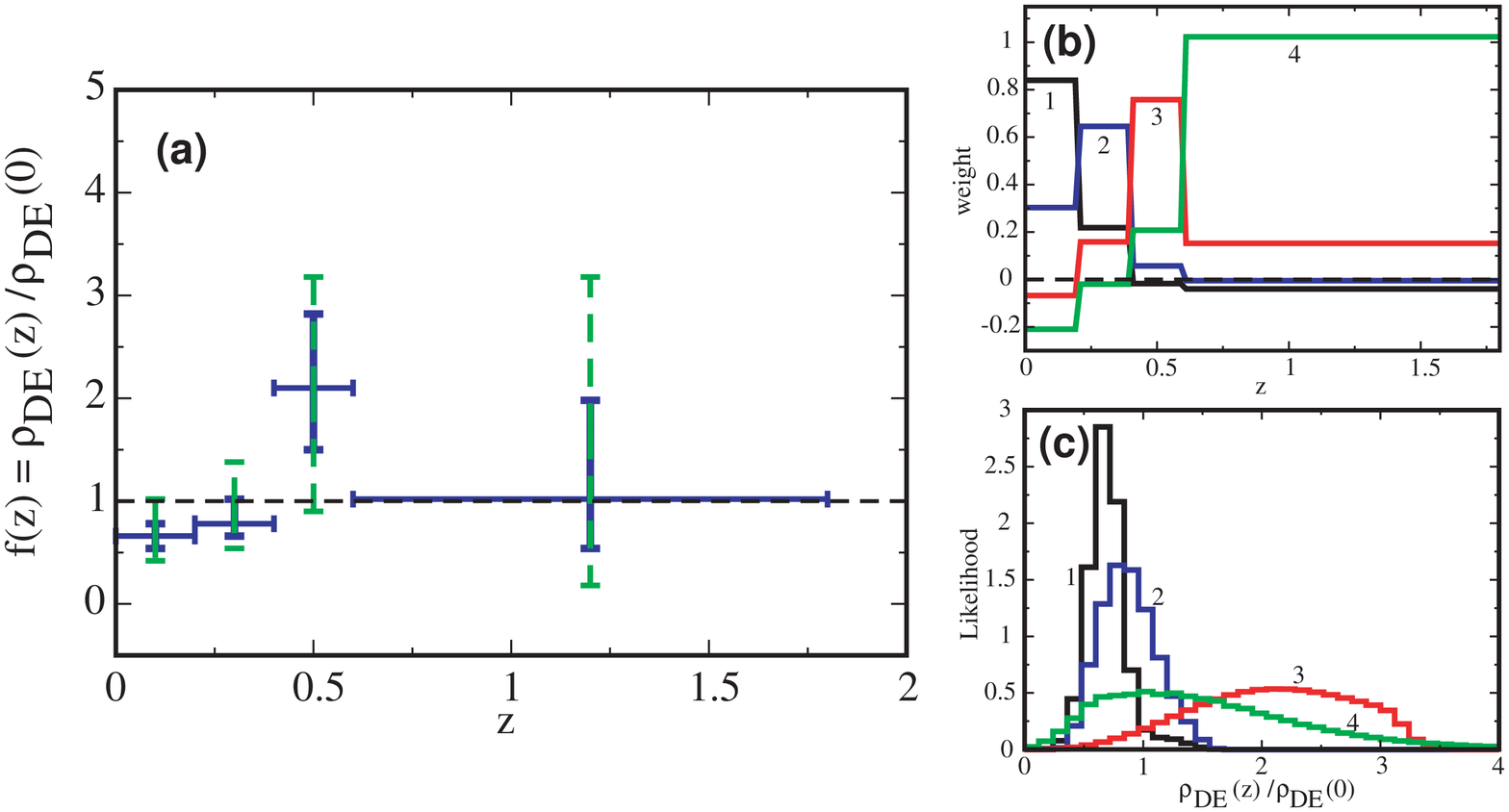,width=6.0in,angle=0}}
\caption{Same as Figure~1, but for $f(z)\equiv \rho_{\rm
DE}(z)/\rho_{\rm DE}(0)$. Our band powers assume piecewise linear
$f(z)$, with the exception of the band power corresponding to
the largest-redshift bin, which assumes constant $f(z)$ across that bin.}
\label{fig:rhoz}
\end{figure*}

\section{Results}\label{sec:results}

We perform the analysis of the ``gold'' dataset from Riess et
al.~\cite{Rieetal04}. First, we need to parameterize $w(z)$ and $f(z)$
in redshift, thereby defining the parameters $p_i$ from
Sec.~\ref{sec:method}. We choose $w(z)$ to be piecewise constant in
redshift and $f(z)$ to be piecewise linear (and continuous). These two
assumptions are consistent, since the two functions are related as
$w(z)=1/3\,(1+z)f'(z)/f(z)-1$. Note that, in the limit of the large
number of parameters $p_i$, the shape of the function across the
redshift bin becomes irrelevant.

We choose $N=4$ bins with redshifts $0\leq z\leq 0.2$, $0.2\leq z\leq
0.4$, $0.4\leq z\leq 0.6$, $0.6\leq z\leq 1.8$, for both $w(z)$ and
$f(z)$ constraints. While our choice of the number of parameters (or
bins) is limited by the computing power required to perform the
maximum likelihood analysis, we have repeated the same analysis with
five parameters in each case and found consistent results.  Future SN
Ia data will lead to better constraints at all redshifts, requiring
more parameters and perhaps the use of Markov chain Monte Carlo
techniques, but for our purpose a simple analysis is sufficient.
Furthermore, we have explored in detail the choice of the redshift
binning, trying to strike a balance between band powers being narrow and
having small error bars. Not surprisingly, we find that the
constraints on $w(z)$ or $f(z)$ are much better at low redshift, and
we put three of our four bins there, choosing their widths so as to
get comparable constraints in each. We have varied the exact spacing
of the bins, and found results consistent with the same underlying
$w(z)$ or $f(z)$.

Finally, we describe the piecewise linear $f(z)$ as follows: we write
$f(z)=1+g(z)$ (note that $g(z)=0$ corresponds to the cosmological
constant scenario).  We describe $g(z)$ by the sawtooth basis in
redshift, where each tooth is $0.2$ wide and peaks in the middle of
the corresponding bin.  The highest-redshift bin presents a problem,
since it is much wider than the others and implies that $f(z)$ may be
forced to vary strongly across this bin. To prevent this, we make all
basis vectors of the sawtooth 100\% correlated in the highest redshift
bin, essentially making $f(z)$ flat across this bin. We have checked
that these details do not affect the results appreciably by repeating
the analysis with a few alternative choices, and we believe that these
assumptions are reasonable and intuitive.

The analysis is now straightforward: we compute the goodness-of-fit
statistic $\chi^2$ for each model in the six-dimensional parameter
space ($p_1\ldots p_4, \Omega_M, h$). We allow a generous range for
the parameters $p_i$ (corresponding roughly to the vertical range in
the left panels of the two Figures) and verify that changing the range
leads to insignificant changes in the final constraints.  We then
marginalize the full likelihood over $\Omega_M$ and all values of $h$
and project it onto the $p_1\ldots p_4$ space. We use a flat prior
$0.22\leq \Omega_M\leq 0.38$, corresponding to the $\pm 2\sigma$
allowed range from the joint analysis of various cosmological probes
\cite{Tegetal03}.  We have repeated our analysis with the Gaussian prior
$\Omega_M=0.30\pm 0.04$ and found that the results are largely
insensitive to the exact choice of either prior: the only notable
change was that the first band power increases by about 0.15 with the
Gaussian prior. The parameters $p_i$ are then rotated into the new
parameters $q_i$, which are now uncorrelated, following the
methodology described in Sec.~\ref{sec:method}.

Figure 1 shows the final 68\% and 95\% CL constraints on the four
band powers (i.e.\ the parameters $q_i$) representing $w(z)$. We also
show the weights that describe going from correlated parameters $p_i$
to the uncorrelated $q_i$, as well as the full likelihoods of the four
band powers. The horizontal error bars in the left panel show the
extent of the original bins; although the components' weights extend
across the whole redshift range, the most weight ($\sim 60$\% or more)
is in these respective bins and the band powers are therefore
sufficiently localized in order to be easily interpreted.  Note also
that the weights are mostly positive and have small negative
contributions, as found in the context of matter power spectrum
measurements~\cite{HamTeg00}.

As shown in Fig.~1, the equation of state is consistent with $w=-1$ at
the 95\% CL in three out of four bins. We do find some ($> 95\%$ CL)
evidence that $w<-1$ at $z< 0.2$; however, to confirm this result with
certainty will require more data, and in particular more stringent
control of the systematic errors.  Nevertheless, it is interesting
that we find a similar tendency in the data as seen in completely
independent analyses that use different, and less general,
parameterizations \cite{Alaetal03, Rieetal04}.  The present approach,
however, is less model dependent than these methods.  In particular,
any variations in the equation of state on redshift scales smoother
than the binning scale can in principle be detected; more rapid
oscillations cannot.  This is why we consider this approach to be {\it
nearly} model independent -- it would be truly model independent if we
used a large number of bins, as illustrated in Ref.~\cite{HutSta03}.

We now consider another parameterization of dark energy -- its energy
density relative to the present value, $f(z)$.  We repeat the analysis
and obtain constraints on $f(z)$ shown in Fig.~2. They are roughly
consistent with those for $w$, and are also consistent with the
cosmological constant case at the 95\% CL. Note that the likelihoods are fully
contained in the allowed ranges, and we see no evidence for negative
$f(z)$.  The weights of $f$ are somewhat less well localized; however,
the band powers are better determined than those of $w$, as expected
from the fact that $f$ is related to the luminosity distance data
through a single, and not double, integral relation.

\section{Discussion and Summary}
\label{sec:Discussion}

We have used a variant of the principal component technique to produce
uncorrelated, nearly model independent estimates of the equation of
state of dark energy $w(z)$ and its scaled energy density $f(z)$.  We
used four redshift bins in each case, and found results that are in
good agreement with previous analyses.  We further argued that the
present approach nicely complements other methods that use
conventional parameterizations of $w(z)$ and $f(z)$. Given that our
band powers are uncorrelated, the interpretation of the cumulative
evidence is particularly easy.

If dark energy is due to the cosmological constant, then $w=-1$ and
$f(z)=1$, and all of our band powers should be consistent with those
values, independently of their window functions. Conversely, if we
ever find strong statistical evidence that {\it even just one}
band power is different from $-1$ (for $w$) or $1$ (for $f$), we will
have ruled out the cosmological constant scenario. While we do find a
hint of such evidence in the first band power of $w(z)$, a definitive
analysis will have to await more data and a careful assesssment of the
systematics.  

While we have presented an analysis with $\sim$ 150 supernovae and
restricted ourselves to four bins in redshift, the generality and
power of this method should make it perfectly suitable
for the analysis of future supernova datasets, when the error bars
are expected to improve by up to an order of magnitude and enable a much more
quantitative analysis and comparison with models. Furthermore, the same techniques
can be applied to a variety of other cosmological probes, as one can
expect that their complementarity will considerably strengthen the SN
Ia results. Finally, one can customize the proposed technique
specifically to maximize the return of any given test (say, whether $w(z)=-1$
or not).  With an increase in the number of type Ia supernovae at high redshift,
it is likely that these interesting possibilities will be considered in the future.

{\it Acknowledgments:} We thank Eric Linder for useful comments on the manuscript. This work has been
supported by the DOE at Case Western Reserve University (DH) and the
Sherman Fairchild foundation and DOE DE-FG 03-92-ER40701 at Caltech
(AC).

\end{document}